\documentclass[10pt, prl, eps,, twocolumn,showpacs,nofootinbib]{revtex4-1}
\usepackage{url,ulem}
\usepackage{fourier}
\usepackage{amssymb,amsmath}
\usepackage[dvipdfmx]{graphicx}
\usepackage{graphicx,color}

\newcommand{\ave}[1]{\left\langle #1 \right\rangle}
\newcommand{\norm}[1]{\left\Vert #1 \right\Vert}
\newcommand{\dfracp}[2]{\dfrac{\partial #1}{\partial #2}}

\newcommand{\bsm}{\boldsymbol{m}}
\newcommand{\bsh}{\boldsymbol{h}}
\newcommand{\bsM}{\boldsymbol{M}}
\newcommand{\bzero}{\boldsymbol{0}}
\newcommand{\bschi}{\boldsymbol{\chi}}

\begin{document}
\title{Strange Scaling and Temporal Evolution of Finite-Size Fluctuation
   in Thermal Equilibrium}
\author{Yoshiyuki Y. Yamaguchi}
\email{yyama@amp.i.kyoto-u.ac.jp}
\affiliation{
  Department of Applied Mathematics and Physics, 
  Graduate School of Informatics, Kyoto University, 
  606-8501, Kyoto, Japan}
\begin{abstract}

  We numerically exhibit strange scaling and temporal evolution
  of finite-size fluctuation in thermal equilibrium
  of a simple long-range interacting system.
  These phenomena are explained from the view point of existence
  of the Casimirs and their nonexactness in finite-size systems,
  where the Casimirs are invariants in the Vlasov dynamics
  describing the long-range systems in the limit of large population.  
  This explanation expects appearance of the reported phenomena
  in a wide class of isolated long-range systems.
  The scaling theory is also discussed as an application
  of the strange scaling.
\end{abstract}

\maketitle 
Studying fluctuation is one of the central issues in large systems,
and physical importance of fluctuation can be found
in Johnson-Nyquist noise \cite{johnson-28,nyquist-28},
fluctuation-induced tunneling \cite{sheng-sichel-gittleman-78,sheng-80},
the fluctuation theorem 
\cite{evans-cohen-morris-93,kurchan-98,jarzynski-00},
and so on.
The fluctuation is also investigated 
in mathematical models of XY rotors on networks \cite{denigris-leoncini-13pre}
and of coupled oscillators \cite{hong-chate-tang-park-15}.
In this article we concentrate on finite-size fluctuation
in isolated Hamiltonian systems having long-range interaction,
which we call long-range systems,
and which include self-gravitating systems,
two-dimensional fluids and plasmas
(see Refs.\cite{campa-dauxois-ruffo-09,levin-pakter-rizzato-teles-benetti-14,campa-dauxois-fanelli-ruffo-14} for instance).
We reveal a strange finite-size scaling
and temporal evolution {\it in thermal equilibrium}.
In order to describe why these phenomena could be possible,
we start from sketching the relaxation process in long-range systems.

A remarkable dynamics in long-range systems
is appearance of the so-called quasistationary state (QSS),
whose life time diverges as population of the system increases
\cite{zanette-montemurro-03,yamaguchi-barre-bouchet-dauxois-ruffo-04,barre-bouchet-dauxois-ruffo-yamaguchi-06}.
The QSSs are widely observed in nature:
Galaxies \cite{binney-tremaine-08}
and the great red spot of Jupiter \cite{bouchet-sommeria-02}
are considered as examples of QSSs.
Temporal evolution of the long-range systems is
governed by the Vlasov equation, or the collisionless Boltzmann equation, 
in the limit of large population
\cite{braun-hepp-77,dobrushin-79,spohn-91},
and QSSs are interpreted as stable stationary solutions
to the Vlasov equation \cite{yamaguchi-barre-bouchet-dauxois-ruffo-04,barre-bouchet-dauxois-ruffo-yamaguchi-06},
which are possibly out-of-equilibrium.
Appearance of QSSs is explained by existence of
infinite number of the Casimirs,
which are invariants of the Vlasov dynamics.
Finite-size effect plays the role of collision,
and the collision term drives the system from a QSS to thermal equilibrium
by breaking the Casimir constraints.

QSSs are observed even applying an external force to the system.
The external force drives the system from an initial state,
for instance a thermal equilibrium state, to a QSS
before going towards a forced thermal equilibrium state.
For the system having a second-order phase transition,
the long life-time of QSSs permits to define the critical exponents
for the response in QSSs, $\gamma_{\pm}$ and $\delta$,
around and at the critical point respectively.
Hereafter, the subscript $+$ (resp. $-$) indicates
that the variable is defined in the disordered (resp. ordered) phase.
A linear \cite{patelli-gupta-nardini-ruffo-12,ogawa-yamaguchi-12}
and a nonlinear \cite{ogawa-yamaguchi-14,ogawa-yamaguchi-15,yamaguchi-ogawa-15}
response theories 
based on the Vlasov description
reveal that $\gamma_{+}=1$
but $\gamma_{-}=1/4$ \cite{ogawa-patelli-yamaguchi-14}
and $\delta=3/2$ \cite{ogawa-yamaguchi-14,ogawa-yamaguchi-15}
in the Hamiltonian mean-field (HMF) model
\cite{inagaki-konishi-93,antoni-ruffo-95},
while equilibrium statistical mechanics gives $\gamma_{\pm}=1$ and $\delta=3$.
The former non-classical critical exponents are again explained
by the Casimir constraints, which suppress the response.

We note that the strange critical exponents for the response
are obtained not in the forced thermal equilibrium but in QSSs.
However, if the fluctuation-response relation holds,
then anomalous fluctuation appears in non-forced {\it thermal equilibrium}.
Similarly, temporal evolution of the finite-size fluctuation can be expected 
since the strange values of $\gamma_{-}$ and $\delta$
comes from the Casimir constraints,
but the finite-size effect releases the system from the constraints
as sketched in the relaxation process.

To examine the above scenario,
we perform direct $N$-body simulations in the HMF model.
We remark that validity of the fluctuation-response relation is not obvious
since the response theories \cite{patelli-gupta-nardini-ruffo-12,ogawa-yamaguchi-12,ogawa-yamaguchi-14,ogawa-yamaguchi-15,yamaguchi-ogawa-15}
are based on the Vlasov dynamics in which the limit of large population is taken.
We, therefore, first confirm the fluctuation-response relation
for off-critical points,
and report the strange scaling at the critical point
corresponding to the non-classical critical exponent $\delta=3/2$.
Then, we show the temporal evolution
of finite-size fluctuation in thermal equilibrium.
We underline that, since the scenario is based on existence of the Casimirs,
the reported phenomena are not limited in the HMF model,
but can be expected in generic long-range systems.
In addition, we discuss that the strange scaling conjectures values of
critical exponents for the correlation length
with the aid of the scaling theory
\cite{fisher-barder-72,botet-jullien-pfeuty-82,botet-jullien-83}.

The HMF model is a paradigmatic long-range system
and is expressed by the Hamiltonian 
\begin{equation}
  \label{eq:HMF}
  H_{N} = \sum_{j=1}^{N} \dfrac{p_{j}^{2}}{2}
  + \dfrac{1}{2N} \sum_{j,k=1}^{N} \left( 1 - \bsm_{j}\cdot\bsm_{k} \right)
  - \bsh\cdot\sum_{j=1}^{N}\bsm_{j}, 
\end{equation}
where $\bsm_{j}=(\cos q_{j},\sin q_{j})$ is a XY spin
and $\bsh=(h_{x},h_{y})$ is the external force.
Each particle is confined on the unit circle,
and the phase of $j$th particle is $q_{j}\in (-\pi,\pi]$
and $p_{j}\in\mathbb{R}$ the conjugate momentum.
The factor $1/N$ in the potential term is added to ensure extensivity of energy.
The magnetization (order parameter) vector of the HMF model is defined by
the arithmetic mean of $\bsm_{j}$ as
\begin{equation}
  \bsM = (M_{x},M_{y}) = \dfrac{1}{N} \sum_{j=1}^{N} \bsm_{j}.
\end{equation}
The HMF model has the second order phase transition
between the disordered phase ($T>T_{\rm c}$)
and the ordered phase ($T<T_{\rm c}$)
with the critical temperature $T_{\rm c}=1/2$ \cite{antoni-ruffo-95}.

We denote the canonical average of $\bsM$ as $\ave{\bsM}$,
and define the zero-field susceptibility tensor $\bschi=(\chi_{ab})$
by statistical mechanics as
\begin{equation}
  \chi_{ab} = \left. \dfracp{\ave{M_{a}}}{h_{b}} \right|_{\bsh=\bzero},
  \quad
  (a,b\in\{x,y\}).
\end{equation}
It is straightforward to prove the relation
\begin{equation}
  \label{eq:fluctuation-response}
  {\rm tr}~\bschi = N \left( \ave{\bsM^{2}} - \ave{\bsM}^{2} \right) / T,
\end{equation}
for $\bsh=\bzero$ from the explicit expression of $\ave{\bsM}$
and the definition of $\bschi$.
We note $\ave{\bsM}=\bzero$ from symmetry of the system.
Further derivations give
\begin{equation}
  \label{eq:fluctuation-Tc}
  \ave{\bsM^{2}} = O(N^{-1/2})
  \quad \text{at } ~ T=T_{\rm c}.
\end{equation}
See Refs.\cite{kittel-shore-65,botet-jullien-83} for instance.

The magnetization vector fluctuates around $\bsM=\bzero$
in the disordered phase.
However, in the ordered phase, the Goldstone mode appears,
and the zero-field susceptibility diverges for this direction
\cite{ogawa-yamaguchi-12,campa-chavanis-10}.
We, therefore in the latter, concentrate on fluctuation
for the radius direction.
Consequently, we compute the finite-size fluctuation by the quantity
\begin{equation}
  \label{eq:VMN}
  V_{M}(N) = \left\{
    \begin{array}{ll}
      \displaystyle{\ave{\bsM^{2}}} & (T\geq T_{\rm c}) \\
      \displaystyle{\ave{\bsM^{2}} - \ave{\norm{\bsM}}^{2}} & (T<T_{\rm c})
    \end{array}
  \right.
\end{equation}
and the fluctuation-response relation is written in the form
\begin{equation}
  \label{eq:fluctuation-response}
  \dfrac{N V_{M}(N)}{T} = \epsilon \chi_{xx},
  \quad
  \epsilon = \left\{
    \begin{array}{ll}
      2 & (T\geq T_{\rm c}) \\
      1 & (T<T_{\rm c}) \\
    \end{array}
  \right.
\end{equation}
from the rotational symmetry,
where the factor $\epsilon$ comes from dimensionality
of the fluctuation of $\bsM$.

We numerically examine
the fluctuation-response relation \eqref{eq:fluctuation-response}
and the fluctuation at the critical point \eqref{eq:fluctuation-Tc}
by direct $N$-body simulations.
We numerically integrate the canonical equations of motion
associated with the Hamiltonian \eqref{eq:HMF}
by using the fourth-order symplectic integrator \cite{yoshida-90}
with the fixed time step $\Delta t=0.1$.
The initial values of $N$ pairs of $(q_{j},p_{j})$ are
randomly picked up from the equilibrium one-particle distribution function
%
$f_{\rm eq}(q,p)=A\exp[-H(q,p)/T]$, where $A$ is the normalization factor,
the one-particle Hamiltonian is $H(q,p)=p^{2}/2-M\cos q$
and $M=\norm{\bsM}$
is the thermal equilibrium value satisfying the self-consistent equation
$M = I_{1}(M/T)/I_{0}(M/T)$
with $I_{n}$ the modified Bessel functions of the first kind.
For $T<T_{\rm c}$, there are the solutions of $M=0$ and $M>0$,
and we use the latter by stability.
We compute $\bsM(t)$ numerically, and replace the canonical averages
in $V_{M}(N)$, \eqref{eq:VMN}, with the time averages defined by
\begin{equation}
  \ave{\bsM^{2}}_{t}
  = \dfrac{1}{t_{\rm av}} \int_{0}^{t_{\rm av}} \bsM^{2}(t) dt,
\end{equation}
for instance.
We further take the average of $V_{M}(N)$ over $100$ realizations
of initial states, but denote it by $V_{M}(N)$ for simplicity.

\begin{figure}
  \centering
  \includegraphics[width=8.5cm]{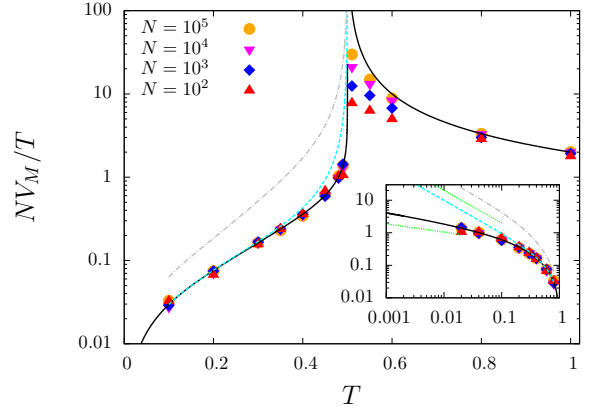}
  \caption{(color online) Comparison between the finite-size fluctuation
    $NV_{M}/T$ (points) and the zero-field susceptibility $\epsilon\chi_{xx}^{\rm V}$
    obtained in the Vlasov dynamics,
    whose level is reported by the black solid lines.
    The dot-dashed gray and the dashed light-blue lines
    in the ordered phase ($T<T_{\rm c}=1/2$) 
    are isothermal and isoentropic susceptibilities
    \cite{ogawa-patelli-yamaguchi-14},
    while they coincide with the Vlasov one in the disordered phase.
    Points are averages over $100$ realizations.
    Averaging time $t_{\rm av}$ for one realization is $t_{\rm av}=100$.
    $N=10^{2}$ (red triangles), $10^{3}$ (blue diamonds),
    $10^{4}$ (magenta inverse-triangles) and $10^{5}$ (orange circles).
    (Inset) Double logarithmic graph in the ordered phase
    with the horizontal axis of the reduced temperature $\tau$.
    Slopes of the two green dotted guide lines are $-1$ and $-1/4$
    from top to bottom.}
  \label{fig:fluctuation-response}
\end{figure}

The finite-size fluctuation $NV_{M}/T$ is compared
with the zero-field susceptibility $\chi_{xx}^{\rm V}$ obtained theoretically
\cite{ogawa-yamaguchi-12} from the Vlasov dynamics 
in Fig.\ref{fig:fluctuation-response}.
They are in good agreement in the ordered phase
and for large $T$ in the disordered phase.
Even around the critical point of the disordered phase,
we can find tendency of convergence of the fluctuation
to the Vlasov susceptibility level as $N$ increases.
Moreover, for the reduced temperature $\tau=|T-T_{\rm c}|/T_{\rm c}$,
the scaling of $NV_{M}/T\propto\tau^{-1/4}$ is found
in the ordered phase corresponding to the non-classical exponent
$\gamma_{-}=1/4$ defined by $\chi_{xx}^{\rm V}\propto\tau^{-\gamma_{-}}$.
Thus, we conclude that the fluctuation-response relation holds
even under the Casimir constraints and the finite-size effects.

\begin{figure}
  \centering
  \includegraphics[width=8.5cm]{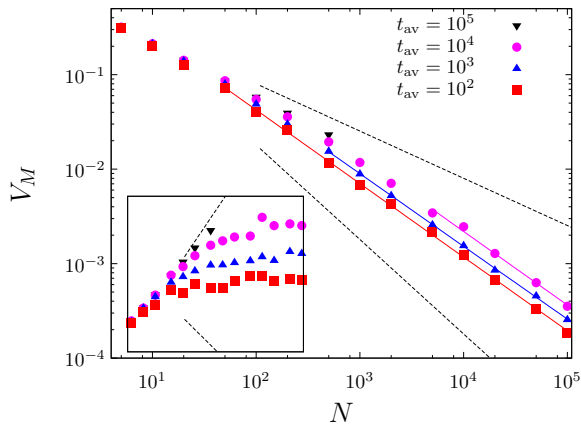}
  \caption{(color online)
    Finite-size fluctuation $V_{M}$ at the critical point $T=T_{\rm c}$.
    Points are averages over $100$ realizations.
    Averaging time $t_{\rm av}$ for one realization are
    $10^{2}$ (red squares), $10^{3}$ (blue triangles),
    $10^{4}$ (magenta circles) and $10^{5}$ (black inverse-triangles).
    The solid lines are computed by the least-mean square method
    in the displayed intervals,
    and their slopes are $-0.779$, $-0.770$ and $-0.779$
    from bottom to top.
    The upper and the lower black dashed lines 
    have the slopes $-1/2$ and $-1$ respectively for comparison.
    (Inset) The vertical axis is $V_{M}N^{4/5}$
    to enhance difference among $t_{\rm av}$.}
  \label{fig:TcM2}
\end{figure}

For the strange scaling at the critical point,
we introduce the Landau's pseudo free energy per one particle
\begin{equation}
  g(M) = a(T-T_{\rm c}) M^{2} + bM^{\delta+1} + \cdots - hM,
  \quad a,b>0.
\end{equation}
In this framework realized $M$ is derived from the equation $dg/dM=0$.
Assuming that the fluctuation level is determined
by the equation $Ng(M)=\Delta$ with a certain value $\Delta$,
we have the scaling of $V_{M}(N)\propto N^{-2/(\delta+1)}$ at the critical point,
and the classical value of $\delta=3$ gives the scaling of \eqref{eq:fluctuation-Tc}.
However, the Vlasov dynamics gives $\delta=3/2$ \cite{ogawa-yamaguchi-14},
and the scaling $V_{M}(N)\propto N^{-4/5}$ is expected accordingly.
This expectation is confirmed in Fig.\ref{fig:TcM2},
and holds even for $N\simeq 100$
with a short time averaging time $t_{\rm av}=100$.
We remark that the scaling tends to approach to the classical one,
$N^{-1/2}$, as increasing $t_{\rm av}$.

\begin{figure}
  \centering
  \includegraphics[width=8.5cm]{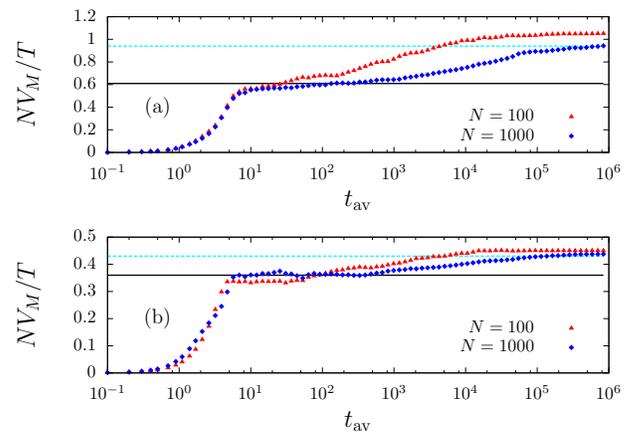}
  \caption{(color online)
    Fluctuation $NV_{M}/T$ as a function of $t_{\rm av}$.
    (a) $T=0.45$. (b) $T=0.40$.
    In both panels
    $N=100$ (red triangles) and $1000$ (blue diamonds).
    The black and the light-blue horizontal lines represent
    levels of the Vlasov and the isoentropic susceptibilities.}
  \label{fig:MdevEvolution}
\end{figure}

As found in Fig.\ref{fig:fluctuation-response},
there is a gap between the Vlasov and the isoentropic susceptibilities,
and hence one may expect that the fluctuation, $NV_{M}(N)/T$,
temporally evolves from the former level to the latter.
This expectation is confirmed in Fig.\ref{fig:MdevEvolution}
for $T=0.45$ and $0.40$ by varying the averaging time $t_{\rm av}$.
The initial increase for $t_{\rm av}\lessapprox 10$
is a natural consequence from the definition
of $V_{M}$ as the time average, and hence we omit it.
The fluctuation is once trapped at the Vlasov level around
$t_{\rm av}=10-1000$ for $N=1000$, and goes towards the isoentropic level.

\begin{figure}[ht]
  \centering
  \includegraphics[width=8.5cm]{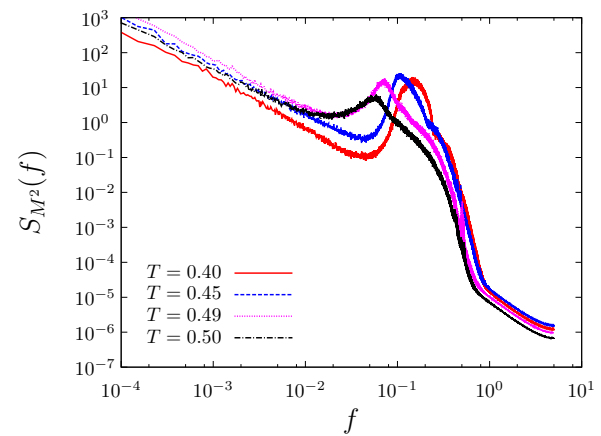}
  \caption{(color online) Power spectra of $\bsM^{2}(t)$
    for $T=0.40$ (red solid), $0.45$ (blue dashed), $0.49$ (magenta dotted)
    and $0.50$ (black dot-dashed) with $N=1000$.
    Power spectra for $T>T_{\rm c}$ are also computed up to $T=0.60$
    (not shown), and they are similar with for $T=0.50$
    except for slight changes of slopes in a small $f$ region.}
  \label{fig:power}
\end{figure}

A qualitative change of dynamics
between a short- and a long-time intervals
is also captured in Fig.\ref{fig:power}
by computing power spectra of $\bsM^{2}(t)$ with $N=1000$.
The power spectra, each of which is the average over $100$ realizations,
are divided into the short-time interval corresponding to $f\gtrsim 0.01$
and the long-time one having algebraic dampings as $f^{-a}$ with
$a$ from $1.0$ to $1.4$.

Putting all together,
we can now make a scenario of the finite-size fluctuation
in the long-range systems.
In the limit $N\to\infty$, the infinite Casimirs
divide the phase space into their level sets,
and any initial state belongs to one of them
and can not escape from the level set.
When $N$ is finite, invariance of the Casimirs is no longer exact
but approximate.
Nevertheless, in the short-time interval,
the system fluctuates along the approximate level set,
which induces the strange scaling of $V_{M}\sim N^{-4/5}$ at the critical point.
As time goes on, the system is released from the approximate constraints,
and the classical scaling $V_{M}\sim N^{-1/2}$ recovers
in the long-time interval.
Temporal evolution of the finite-size fluctuation can be understood
as the releasing process from the iso-Casimir contour.
An important remark is that any initial states must evolve
under the approximate but unavoidable Casimir constraints
even in thermal equilibrium.

\begin{figure}[h]
  \centering
  \includegraphics[width=8.5cm]{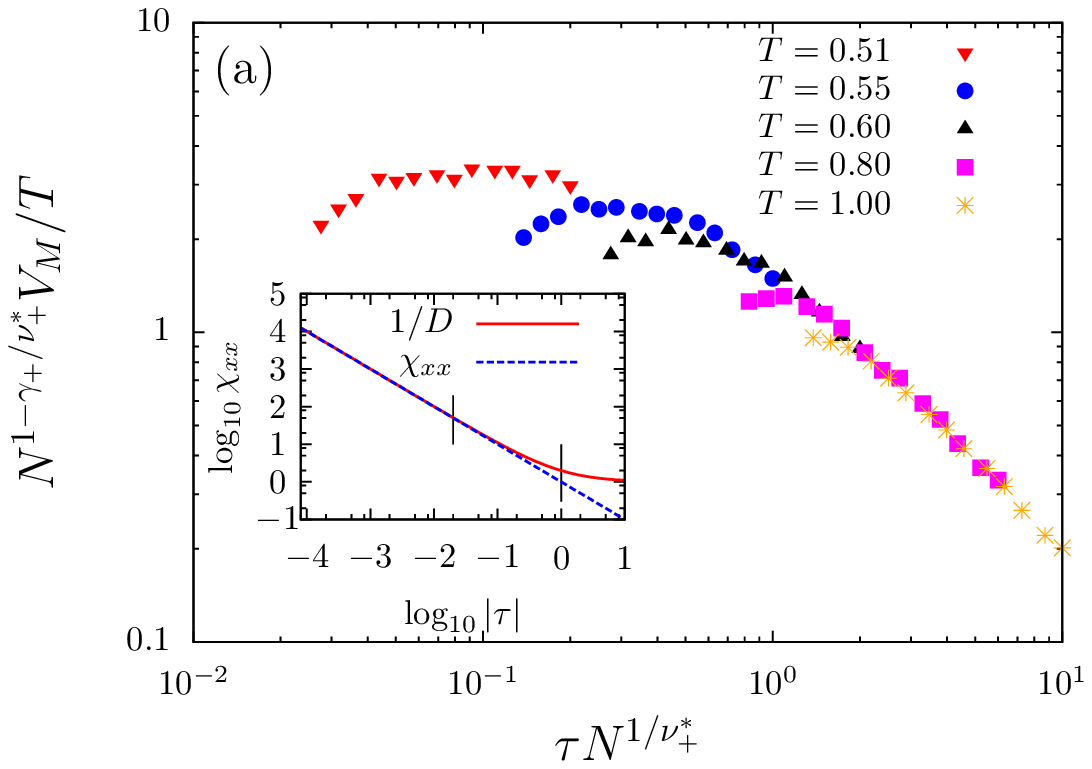}
  \includegraphics[width=8.5cm]{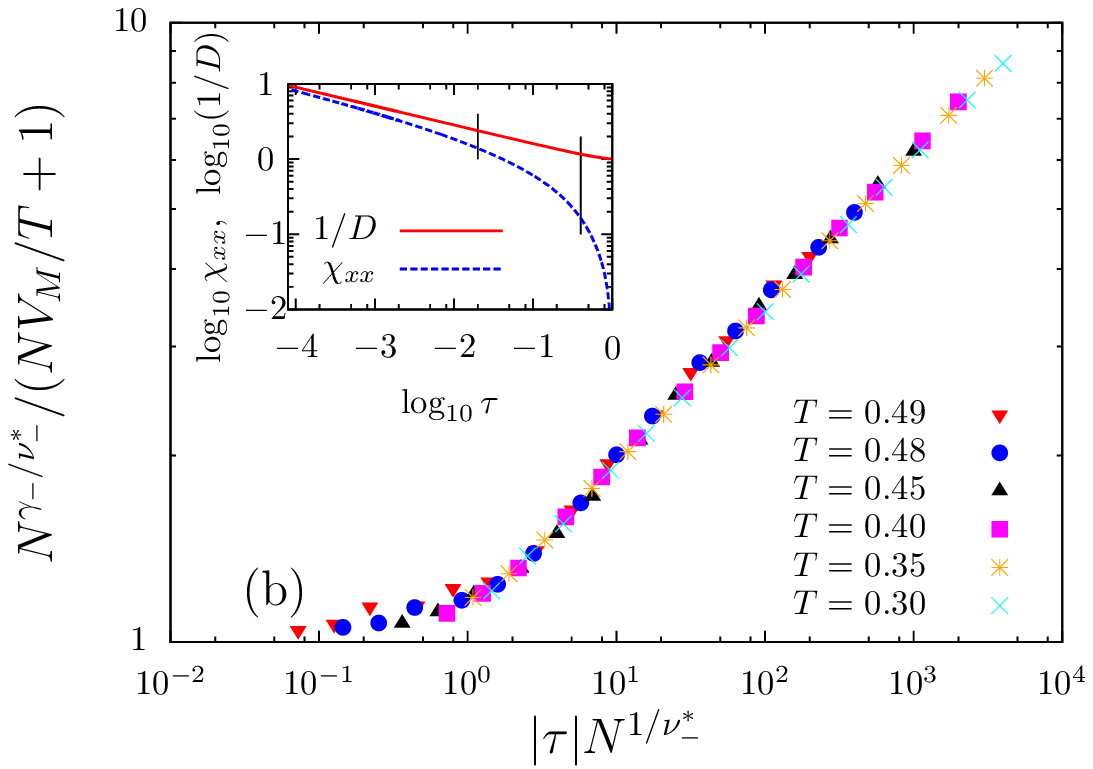}
  \caption{(color online)
    Examinations of the scaling theory.
    (a) Disordered phase. (b) Ordered phase.
    (Insets) $\chi_{xx}$ for the Vlasov dynamics (blue dashed lower lines)
    and $1/D$ (red solid upper lines).
    The vertical black segments mark the computing interval of $\tau$.}
  \label{fig:scaling}
\end{figure}

We further investigate the finite-size fluctuation
from the view point of the scaling theory \cite{fisher-barder-72}.
One hypothesis of the scaling theory is
that the finite-size scalings are controlled by the dimensionless
quantity $\xi/L$, where $\xi$ and $L$ are the correlation length
and the system length respectively.
The infinite-range models, including the HMF model,
have no concept of both lengths,
but a scaling theory has been proposed
for such models by replacing $\xi/L$ with $N_{\rm c}/N$
\cite{botet-jullien-pfeuty-82,botet-jullien-83},
where $N_{\rm c}$ is the coherent number of particles.
It is supposed that 
$N_{\rm c}\propto \xi^{d_{u}} \propto \tau^{-d_{u}\nu}$
with $d_{u}$ the upper critical dimensionality
and $\nu$ defined by $\xi\propto\tau^{-\nu}$
for the mean-field universality class.
Thus, $\nu^{\ast}=d_{u}\nu$ plays the role of $\nu$,
and the scaling function for $\chi_{xx}$, denoted by $F_{\chi}$,
can be introduced as
\begin{equation}
  \label{eq:scaling-function-chi}
  \chi_{xx} = N^{\gamma/\nu^{\ast}} F_{\chi}(\tau N^{1/\nu^{\ast}}).
\end{equation}
Combining the above expression with the fluctuation-response relation
\eqref{eq:fluctuation-response}, we have
\begin{equation}
  \label{eq:scaling-theory}
  N^{1-\gamma/\nu^{\ast}}V_{M}(N)/T
  = \epsilon F_{\chi}(\tau N^{1/\nu^{\ast}}).
\end{equation}
The scaling \eqref{eq:scaling-theory} is examined
in Fig.\ref{fig:scaling}(a) for the disordered phase,
and the scaling function $F_{\chi}$ can be observed.
In the ordered phase, the susceptibility $\chi_{xx}$
scales as $\chi_{xx}\propto \tau^{-\gamma_{-}}$
but this scaling is not excellent even for rather small $\tau$,
and observation of the scaling function is hard accordingly.
The susceptibility is represented as
$\chi_{xx}=(1-D)/D$, and $D$ shows a beautiful scaling
as shown in the inset of Fig.\ref{fig:scaling}(b).
We, therefore, examine the scaling for $D$, which suggests
\begin{equation}
  \dfrac{N^{\gamma_{-}/\nu^{\ast}_{-}}}{NV_{M}(N)/T+1}
  =  F_{D}(\tau N^{1/\nu^{\ast}_{-}}).
\end{equation}
The scaling function $F_{D}$ is clearly observed
in Fig.\ref{fig:scaling}(b),
and hence we may also expect the scaling function $F_{\chi}$
in the ordered phase for small values of $\tau$.

The scaling theory suggests the values of 
critical exponents $\nu_{+}$ and $\nu_{-}$
for the correlation length.
We showed that $V_{M}(N)\propto N^{-4/5}$ at the critical point $\tau=0$,
and hence the relation \eqref{eq:scaling-theory} determines
$\nu_{+}^{\ast}=5$ and $\nu_{-}^{\ast}=5/4$
from $\gamma_{+}=1$ and $\gamma_{-}=1/4$
\cite{ogawa-yamaguchi-12} respectively.
Thus, inputting $d_{u}=4$, $\nu_{+}=5/4$ and $\nu_{-}=5/16$
can be conjectured in QSSs.
These exponents should be tested
in the $\alpha$-HMF model \cite{anteneodo-tsallis-98} for instance,
in which interaction depends on the distance between a pair of 
interacting particles.

Summarizing, we investigated the finite-size fluctuation
in thermal equilibrium of the HMF model.
We have confirmed that the fluctuation-response relation holds
even under the Casimir constraints and finite-size effects.
This relation induces the strange scaling of $V_{M}(N)\propto N^{-4/5}$
at the critical point instead of $N^{-1/2}$ predicted by statistical mechanics.
Further, we reported temporal evolution of the finite-size fluctuation.
These phenomena are explained by existence of the Casimirs,
which are approximate invariants in finite-size long-range systems.
Therefore, these phenomena can be expected in generic systems.
In addition, the strange scaling with the scaling theory conjectures
critical exponents for the correlation length.

We end this article by giving five remarks.
First, the linear and the nonlinear response theories
are simply reproduced \cite{ogawa-yamaguchi-15}
by the so-called rearrangement formula \cite{yamaguchi-ogawa-15}.
It is worth extending the powerful formula for finite-size systems.
Second, the evolution of finite-size fluctuation is not expected
in the disordered phase, since there is no gap 
between the Vlasov susceptibility and the isoentropic one.
Third, nonexactness of the Casimir invariants is a crucial point
for the evolution. Exact invariants
of the translational and the angular momenta break
the equipartition of kinetic energy in small clusters,
but this phenomenon is understood by inputting the invariants
into statistical mechanics \cite{niiyama-07}.
Fourth, approximate constraints can be found also in spring-chain systems.
Hard springs play the role of approximate constraints,
lengths of bonds, and slow relaxation to equipartition is observed
\cite{konishi-yanagita-10}.
The linked particles can be considered as models of molecules
for instance, thus it might be interesting to investigate
the finite-size fluctuation in such models.
Finally, the strange scaling is observed for a rather small $N$,
say $N\simeq 100$, when the averaging time is short.
This observation suggests that we need to consider
the wreck of Casimirs even in small systems,
and might give a new perspective to understand more realistic systems.

\acknowledgements
The author thanks K. A. Takeuchi for suggesting
to study the finite-size fluctuation.
He also acknowledges the support of JSPS KAKENHI Grant Number 23560069.

\end{document}